\documentclass[sigchi-a]{acmart}

\def\BibTeX{{\rm B\kern-.05em{\sc i\kern-.025em b}\kern-.08emT\kern-.1667em\lower.7ex\hbox{E}\kern-.125emX}}
    
\copyrightyear{2019}
\acmYear{2019}
\setcopyright{none}
\acmConference[HCMLP'19]{Human-Centered Machine Learning Perspectives Workshop}{May 04, 2019}{Glasgow, UK}
\acmBooktitle{Human-Centered Machine Learning Perspectives Workshop, May 04, 2019, Glasgow, UK}

\begin{document}

%
\title{An Interaction Framework for Studying Co-Creative AI}

%
\author{Matthew Guzdial}
\email{mguzdial3@gatech.edu}
\affiliation{%
  \institution{Georgia Institute of Technology}
  \city{Atlanta}
  \state{Georgia}
}

\author{Mark Riedl}
\email{riedl@cc.gatech.edu}
\affiliation{%
  \institution{Georgia Institute of Technology}
  \city{Atlanta}
  \state{Georgia}
}

%
\renewcommand{\shortauthors}{Guzdial and Riedl}

%
\begin{abstract}
Machine learning has been applied to a number of creative, design-oriented tasks. 
However, it remains unclear how to best empower human users with these machine learning approaches, particularly those users without technical expertise.
In this paper we propose a general framework for turn-based interaction between human users and AI agents designed to support human creativity, called {co-creative systems}. 
The framework can be used to better understand the space of possible designs of co-creative systems and reveal future research directions. 
We demonstrate how to apply this framework in conjunction with a pair of recent human subject studies, comparing between the four human-AI systems employed in these studies and generating hypotheses towards future studies.
\end{abstract}

%
%


%

%

%
\maketitle

\section{Introduction}

In recent years there have been many machine learning (ML) systems that demonstrate the ability to create in domains like storytelling, music, visual art, and games \cite{martin2018event,saito2018statistical,gatys2015neural,summerville2017procedural}.
Modern ML systems have a high technical knowledge requirement. 
Making creative artifacts with ML often requires directly tweaking a system's parameters, which does not fit naturally into creative workflows \cite{joshi2017bringing}.
{\em Co-creative systems}, also called mixed-initiative \cite{allen1999mixed} or centaur \cite{goldstein2017human} systems, serve as an alternative paradigm to fully autonomous systems by allowing a human creator to work alongside an AI algorithm toward a joint goal.
Co-creative systems can help to democratize creative ML algorithms and support human creativity.
However, the question of how to design the interaction between ML algorithm and human creator remains under-explored.

There have been several prior attempts to classify the kinds of interaction paradigms computers and humans might engage in. 
Three common classes are: (1) creativity support tool, (2) autonomous, generative system, and (3) co-creative system \cite{davis2015enactive,karimi2018evaluating}. 
Lubart \cite{lubart2005can} suggested that computers can play four roles in creativity support: {\em nanny}, {\em pen-pal}, {\em coach}, and {\em colleague}. 
These prior frameworks or classifications are  under-specified, making it difficult to apply them to meaningfully compare between systems for the purpose of evaluation or hypothesis generation. 

In this paper we present a framework for representing turn-based, co-creative systems.
In particular, we present structure and language for identifying and discussing the common components of these systems across domains.
we demonstrate how this framework can be used to represent Lubart's four roles, and the three common classes described above.
We further present the utility of this framework for eliciting new research hypotheses about co-creative systems in a case study based on prior human subject studies and simulated tests.
This framework is designed for comparing existing systems, as a framing device during initial design of novel systems, and as a tool for generating hypotheses.

\section{Framework}

\begin{marginfigure}
	\includegraphics[height=5.5in]{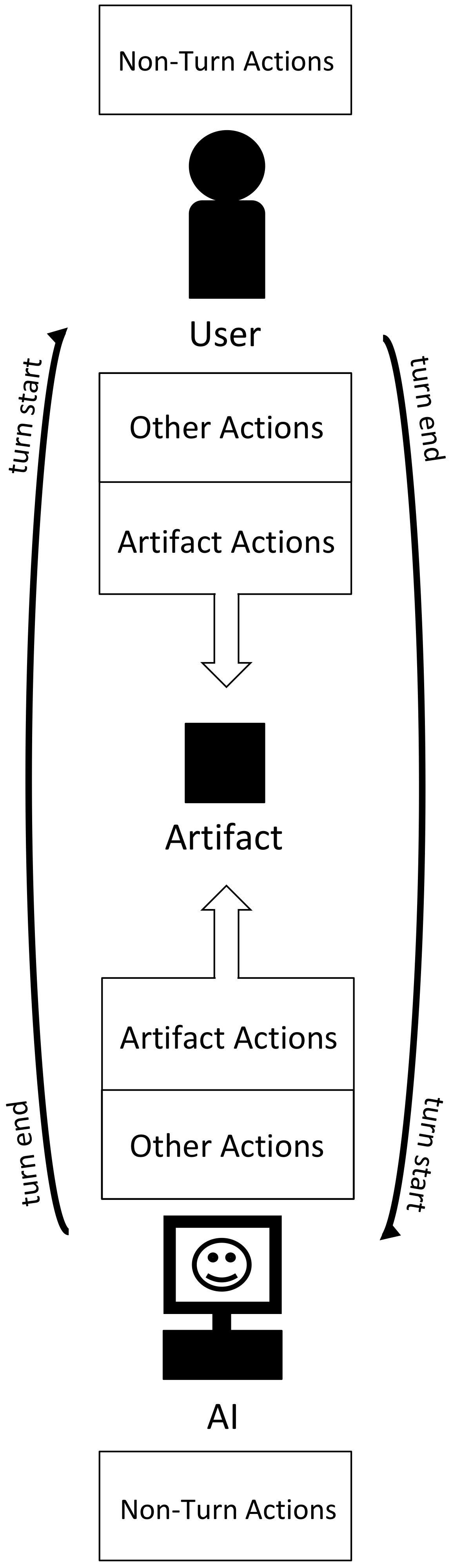}
	\caption{Illustration of our framework.}
	\label{fig:framework}
\end{marginfigure}

We present our framework in Figure \ref{fig:framework}. 
It describes systems with a human \textbf{user} and \textbf{AI} agent collaboratively working to design some \textbf{artifact} in a turn-based fashion. We employ a turn-based framework because it is a common way of organizing co-creative interactions~\cite{liapis2013designer,davis2015enactive,karimi2018evaluating} and because it suits evolutionary and reinforcement-learning approaches that require discrete steps~\cite{liapis2013sentient,alvarez2018fostering,guzdial2018co,guzdial2019friend}. 

Each component of this framework is meant as both a representation of some part of a system and a means of prompting design reflection:
\begin{enumerate}
  \item \textbf{Start}: How does the interaction start? Who has the first turn?
  \item \textbf{Actions}: What actions is each partner capable of based on implementation and interface design.
  \begin{enumerate}
     \item \textbf{Artifact Actions}: During each turn, what are the actions each partner can take during their turn that directly impact the artifact?
     \item \textbf{Other Actions}: During each turn, what are the actions that do not directly impact the artifact? For example, an AI agent explaining its actions or internal state. \cite{zhuexplainable,guzdial2018explainable}.
     \item \textbf{Non-Turn Actions: } What actions can each partner take outside of their turn? For example does the AI attempt to actively learn from or model the user. \cite{liapis2013designer,guzdial2019friend}? Does the user have the ability to observe or explore the artifact?
    \end{enumerate}
  \item \textbf{AI}: What algorithm or technique is used to train or design the AI agent? Is this agent static, choosing between a series of existing behaviors, or learning during the interaction?
  \item \textbf{User}: Who is the imagined or intended user for this system? Prior work has demonstrated the importance of designing these systems towards a particular target user \cite{kantosalo2014isolation}. 
  \item \textbf{Turns}: How do turns begin and end? Which partner has agency over these decisions?
  \item \textbf{End}: How does the interaction end? Which partner or conditions must be satisfied?
\end{enumerate}

\subsection{Comparison to Prior Frameworks}

Three common ways of describing interaction between an AI agent and a human user are: (1) creativity support tool, (2) autonomous, generative system, and (3) co-creative system. All three of these can be expressed with our proposed framework. 
First, a creativity support tool can be represented in this framework with the AI agent having no Artifact Actions, only taking actions when the user is not.
Second, an autonomous generative system instead involves the user not having any Artifact Actions, but a single other action that starts the AI process that completes the artifact, thus ending the AI agent's only turn.
The third involves a human and an AI agent taking turns and most clearly fits this framework.
Still, there are many ways in which humans and AI agents can co-create and our framework allows for a more fine-grained level of description.

Lubart \cite{lubart2005can} presents an influential framework that classified computer-human interactions into four categories: {\em nanny}, {\em pen-pal}, {\em coach}, and {\em colleague}. 
A computer-as-nanny system conducts regular check-ins to monitor, support, or manage the user.
Our framework can represent nanny-like interactions as Other Actions and Non-turn Actions taken by an AI agent that has no Artifact Actions.
A computer-as-pen-pal system sends and receives messages from the human creator in order to encourage creative reflection.
Our framework represents this by removing Artifact Actions from both user and AI agent.
A computer-as-coach system teaches the user strategies and skills for creation.
Our framework can represent a coach with an empty set of Artifact Actions and using Other Actions and Non-turn Actions that teach and provide constructive feedback.
A computer-as-colleague system acts as a peer, taking turns to contribute to the artifact.
In our representation, this is the only interaction configuration in which the AI agent can make Artifact Actions. 
A computer-as-colleague system can also have Other Actions and Non-turn Actions, providing a wide range of possible agent interaction designs.

\subsection{Case Study}

\begin{marginfigure}
	\includegraphics[width=3in]{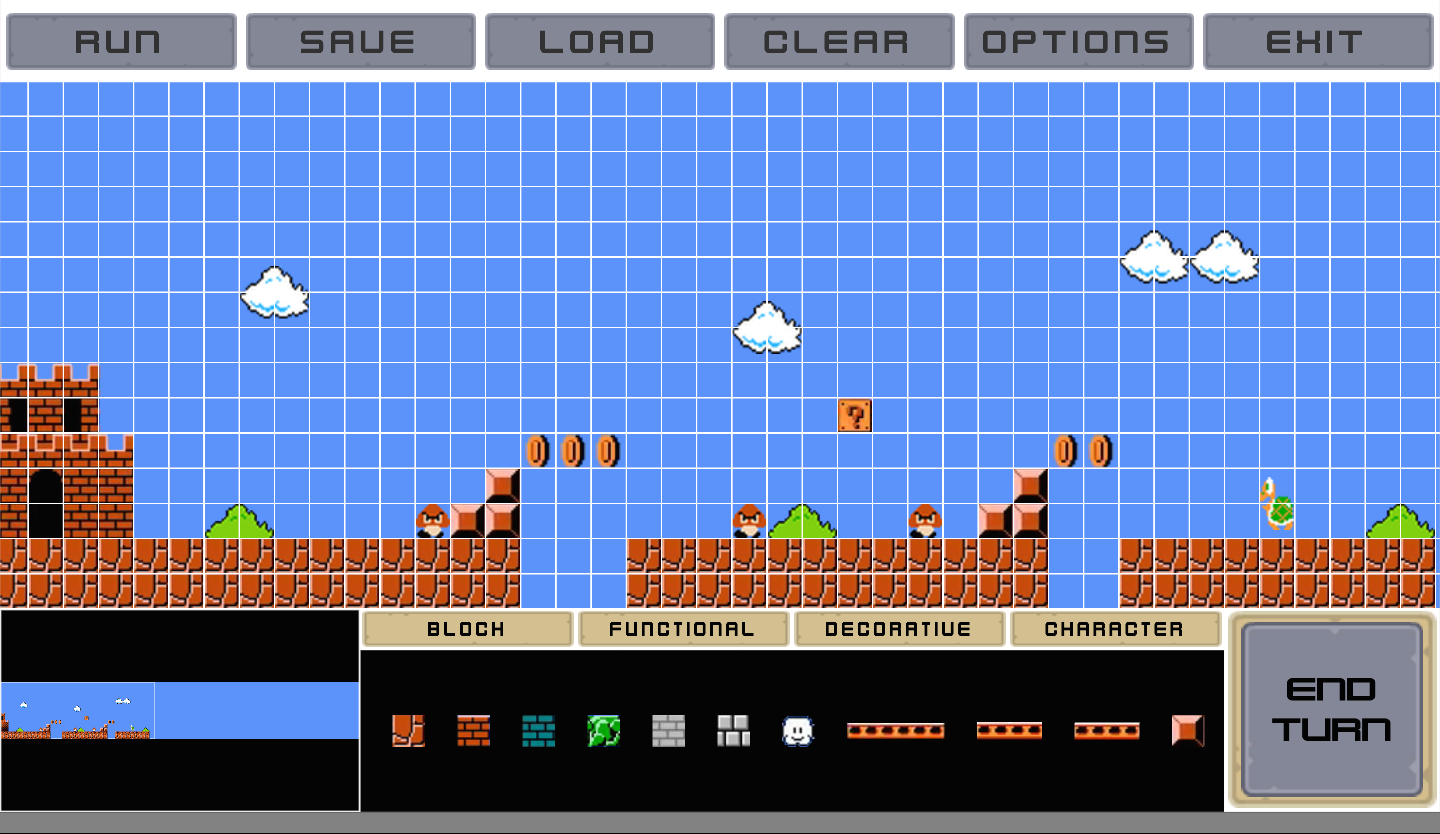}
	\caption{Screenshot of the interface for the systems in the first study.}
	\label{fig:screenshot}
\end{marginfigure}

Our co-creative agent framework allows us to draw comparisons between any two systems that can be represented within it. 
In this section we discuss our recent work on studying co-creativity in the context of a \textit{Super Mario Bros.} level design editor.
Two studies, described in~\cite{guzdial2019friend}, sought to qualitatively understand how users of co-creative systems thought of the AI agent and compared different ML-based AI partners. The first study interface is shown in Figure \ref{fig:screenshot} and allowed users to interact with one of three distinct ML-based AI partners in the backend: an LSTM, a Markov Chain, and a Bayes Net.\footnote{A video of the Bayes Net agent can be found here: https://youtu.be/6Ska6Y9Wnvo} 
Between the studies we developed the interface further, removing an ability for the user to play their current level during their turn, added the ability to ask a study conductor for an explanation of the AI agent's last action (simulating a future explainable AI sub-system), and added a button corresponding to a user action undoing all of the AI agent's last turn. 
We also replaced the AI agent with an active learning, deep reinforcement agent.\footnote{Video of this agent can be found here: https://youtu.be/UkMeM5Ty1lA}
We note that this paper's framework is not described in~\cite{guzdial2019friend}.

For an example of how our framework helps guide the study of co-creative systems, we can describe the impact the AI algorithm (Markov Chain, Bayes Net, and LSTM) had on the user's perceptions of the co-creative system.
We found surprisingly few statistically significant differences in user perception across these systems. 
There was a significant results found that the Bayes Net was more creative in behavior than the LSTM, that the LSTM was more challenging to use than the Bayes Net, and that the Markov Chain was more creative in behavior than the Bayes Net.
Represented in the language of our framework we anticipate this was due to the Bayes Net producing less changes per action compared to the LSTM or Markov Chain, while also including actions to add decorative elements that both other agents did not.
All three agents had an ``Other Action'' for observing the current level, which differed in scale. The Markov Chain perceived only small windows of the current level, the Bayes Net a full screen, and the LSTM almost the entire level. 
This influenced the results we found, with the Markov Chain adding more locally interesting content, leading to the higher perception of its creativity.

Changing the other actions available to the user changed their experience working with the co-creative system.
In the second study using a Deep RL agent, we added an undo button. 
Surprisingly, users kept an average of 15\% more of the AI agent's changes compared to the three systems without an undo button and self-reported 15\% less frustration over all. 
While the latter seems to follow from the addition of the undo button, the former goes against expectations.
However, this undo button could certainly have helped the Deep RL AI agent's non-turn action of actively learning from the user, giving a very clear signal when its changes weren't appreciated and allowing it to learn faster.

Including active learning as a non-turn action for the Deep RL AI agent improved simulated test performance \cite{guzdial2018co} and human self-reports \cite{guzdial2019friend}. 
In the second study we were surprised to find that despite designing the Deep RL System with Lubart's computer-as-colleague interaction framework in mind, users perceived the agent as a friend, collaborator, student, and manager \cite{guzdial2019friend}.
This suggests that even with an otherwise fixed system, variances in users can lead to different perceptions of the AI partner.

\subsection{Research Hypothesis Generation}


The above case study is not exhaustive in terms of research questions about how to design more effective co-creative systems. 
The framework suggests that there may be human-subject studies involving other parts of the framework that the case study didn't address.
For example, the framework tells us that research questions can be crafted around how turns are ended.
Anecdotal evidence suggests that users that interacted with the AI more had a more positive experience. 
Could the frequency of turns be altered to further increase positive perceptions of the creative experience, perhaps through predictive interruptions?
Likewise, should the user be able to interrupt the AI agent?

There are numerous types of other actions that can be considered where the user interacts with or explores what they have created without directly changing the artifact. 
For example, testing or any means of deriving additional observations of the artifact would be considered an other action.

What can and should the user be doing while the AI agent is taking its turn to increase engagement?
Can the user request explanations for what the AI agent is doing when changing the artifact? Explanations have been shown to positively affect user perceptions of intelligent systems \cite{guzdial2018explainable,ehsan2019automated}. 
Should explanations be offered automatically as an Other Action belonging to the AI agent?

\section{Conclusions}

In this paper we present a general framework for representing turn-based co-creative systems. 
Such a framework allows for direct comparisons between distinct systems, from which one can generate discussion, hypotheses, and potential generalizations on the impact of these differences. We anticipate that this framework can further serve as a means of eliciting design reflection during the development stages of such a tool. Organizing frameworks like this can serve as a vector for developing general design knowledge for best practices in developing usable machine learning approaches.

%
\begin{acks}
This material is based upon work supported by the National Science Foundation under Grant No. IIS-1525967. Any opinions, findings, and conclusions or recommendations expressed in this material are those of the author(s) and do not necessarily reflect the views of the National Science Foundation.
\end{acks}

%
\bibliographystyle{ACM-Reference-Format}
\bibliography{sample-sigchi-a}

\end{document}